\documentclass{jfm}
\usepackage{graphicx}
\usepackage{epstopdf, epsfig}
\usepackage{color}
\usepackage{subfigure}
\usepackage{amsmath}
\usepackage{multirow}

\title{Two-dimensional energy spectra in high Reynolds number turbulent boundary layers}
\author{Dileep Chandran
	\corresp{\email{dpadinjare@student.unimelb.edu.au}},
	Rio Baidya,
	Jason P. Monty
	\and Ivan Marusic}

\affiliation{ Department of Mechanical Engineering, The University of Melbourne,
	Victoria 3010, Australia}

\begin{document}
\allowdisplaybreaks
\maketitle

\begin{abstract}
	Here we report the measurements of two-dimensional (2-D) spectra of the streamwise velocity ($u$) in a high Reynolds number turbulent boundary layer. A novel experiment employing multiple hot-wire probes was carried out at friction Reynolds numbers ranging from 2400 to 26000. Taylor's frozen turbulence hypothesis is used to convert temporal-spanwise information into a 2-D spatial spectrum which shows the contribution of streamwise ($\lambda_x$) and spanwise ($\lambda_y$) length scales to the streamwise variance at a given wall height ($z$). At low Reynolds numbers, the shape of the 2-D spectra at a constant energy level shows $\lambda_y/z \sim (\lambda_x/z)^{1/2}$ behaviour at larger scales, which is in agreement with the existing literature at a matched Reynolds number obtained from direct numerical simulations. However, at high Reynolds numbers, it is observed that the square-root relationship tends towards a linear relationship  ($\lambda_y \sim \lambda_x$) as required for self-similarity and predicted by the attached eddy hypothesis.
\end{abstract}

\begin{keywords}
	turbulent boundary layers, boundary layer structure	
\end{keywords}

\vspace{-10mm}
\section{Introduction}
A number of models for the inertial (logarithmic) region of wall-turbulence rely on the assumption that the coherent and energy-containing motions (or eddies) are self-similar. This self-similarity requires that the geometric length and velocity of the turbulent structures scale with wall height ($z$) and friction velocity ($U_\tau$)  respectively. \citeauthor{townsend1980structure}'s (\citeyear{townsend1980structure}) attached eddy hypothesis is arguably the best known of these models and further assumes a random arrangement of the self-similar eddies that are attached to the wall and have a population density that is inversely proportional to their size. Townsend's conceptual model was extended by \citet{perry1982mechanism}, who specified shapes of eddies to make more detailed predictions. Further, on the basis of dimensional analysis, Perry \& Chong predicted a $k_x^{-1}$ scaling of the one-dimensional (1-D) streamwise spectra of $u$ where $k_x$ is the streamwise wavenumber. The $k_x^{-1}$ scaling, which later gained support through spectral overlap arguments of \citet{perry1986theoretical} and is commonly assumed in atmospheric boundary layer research \citep{hunt2000eddy}, suggests the existence of a range of length scales that contribute equally to the turbulent kinetic energy. However, $k_x^{-1}$ scaling of the 1-D streamwise spectra as an indicator of self-similarity in turbulent boundary layers has remained elusive even in high Reynolds number experiments \citep{nickels2005evidence,rosenberg2013turbulence}. While the lack of empirical support questions the existence of such a scaling, \citet{davidson2006logarithmic} argue that 1-D energy spectra are not the ideal tools to observe self-similarity. Their work showed that aliasing contaminates the 1-D streamwise spectra by artificially shifting the energy to lower wavenumbers, thereby distorting the spectrum and hiding the extent of $k_x^{-1}$ spectral scaling.

The aliasing problem arises because a 1-D spectrum is measured along a straight line and the wave number vectors inclined to the line of measurement could be misinterpreted as low wave number disturbances \citep{tennekes1972first}. That is, a streamwise 1-D spectrum lacks directional information along the spanwise direction and is only the average energy contribution over the entire range of spanwise wavelengths, $\lambda_y$ ($ = 2\pi/k_y$), for a particular streamwise wavelength, $\lambda_x$ ($ = 2\pi/k_x$). Direct measurement of the 2-D spectrum, which is a function of both $k_x$ and $k_y$, avoids the aliasing problem (at least in the homogeneous directions). However, analysis of the 2-D spectrum of wall-turbulence is rare. One of the earliest experiments was carried out by \cite{morrison1969structural} using hot-wire anemometry and investigated 2-D frequency-transverse wave number spectra in narrow frequency bands to report structural self-similarity in the logarithmic region of a fully developed pipe flow. In turbulent boundary layers, 2-D spectral information at low to moderate Reynolds numbers were measured by \cite{tomkins2005energetic} using PIV. They studied the average growth of energetic motions throughout the logarithmic region by considering the energy containing spanwise modes for various streamwise length scales. A dimensional consideration of the 2-D spectrum in the inertial region was made by \citet{chung2015universality}, who argued that in order to have a $k_x^{-1}$ behaviour in the 1-D spectrum, a region of constant energy in the 2-D spectrum should be bounded by  $\lambda_y/z \sim f_1(\lambda_x/z)$ and $\lambda_y/z \sim f_2(\lambda_x/z)$ where $f_1$ and $f_2$ are identical power laws.  Further, \citet{del2004scaling} examined 2-D spectra from direct numerical simulations (DNS) of channel flow at $Re_\tau \leq 1900$ and showed that such a region of constant energy is bounded at larger scales by a square-root relationship of the form 
$\lambda_y/z \sim (\lambda_x/z)^{1/2}$. 
(Here, the friction Reynolds number is given by $Re_\tau = \delta U_\tau/\nu$, where $\delta$ is boundary layer thickness or channel half-height, $U_\tau$ is friction velocity and $\nu$ is kinematic viscosity.) 
This square-root relation between the lengths and widths of the structures results in a failure of self-similarity. In contrast, the attached eddy hypothesis, with its geometrically self-similar eddies, requires the linear relationship $\lambda_y \sim \lambda_x$. However, it is noted that Townsend's arguments are only expected to strictly hold at high Reynolds number. Therefore to validate the assumption of self-similar eddies in the attached eddy model, 2-D spectra at high Reynolds numbers are required. 
In the following we present experimental measurements of 2-D spectra over a range of Reynolds numbers, up to $Re_\tau \approx 26000$.

It is noted that throughout this paper superscript `$+$' indicates the normalization using viscous length and velocity scales, which are $\nu /U_\tau$ and $U_\tau$ respectively.  The streamwise, spanwise and wall-normal directions are denoted by $x$, $y$ and $z$ respectively, and $u$ denotes the streamwise fluctuating velocity component. 

\section{Experimental set-up and methods}
Two sets of experiments were conducted in zero-pressure-gradient boundary layers, in two separate facilities. The first, which we refer to as the low Reynolds number experiments were conducted in an open return boundary layer wind tunnel as described by \cite{monty2011parametric}, and the high Reynolds number experiments were conducted in the large Melbourne wind tunnel (HRNBLWT; \citealp{baars2016wall}). Details of the experiments are summarized in table \ref{table:exper}. Here, the boundary layer thickness ($\delta$) is defined by fitting the velocity profile to the composite profile of \citet{chauhan2009criteria}, while $l$, $U_\infty$ and $T$ denote the hot-wire sensor length, freestream velocity and total sampling duration. 
\begin{table}
	\centering
	\begin{tabular}{c c c c c c c c c c c c c}
		\multicolumn{6}{c}{Low Re facility} & &\multicolumn{6}{c}{HRNBLWT} \\
		\multicolumn{6}{c}{\rule{6cm}{0.5pt}}  & & \multicolumn{6}{c}{\rule{6cm}{0.5pt}}\\	 
		$Re_\tau$ & $U_\infty$ & $\delta$   &  $z^+$ & $l^+$ & $TU_\infty/\delta$ & &$Re_\tau$ & $U_\infty$ & $\delta$   &  $z^+$ & $l^+$ & $TU_\infty/\delta$\\
		& (m/s) & (m)  & &  &  ($\times 10^3$)  & & & (m/s) & (m)  & &  & ($\times 10^3$)\\ 
		\multicolumn{6}{c}{\rule{6cm}{0.4pt}}  & & \multicolumn{6}{c}{\rule{6cm}{0.4pt}}\\	 
		2430  & 15 & 0.056 &  {\color{red} 116}  &17  &32 & &15100 & 20 & 0.37  &  150 &20  & 19.5 \\
		2430  & 15 & 0.056 &  150 		&17  &32 & &15100 & 20 & 0.37  &  {\color{red} 320}  &20  & 19.5\\
		2430  & 15 & 0.056 &  200 		&17  &32 & &20250 & 30 & 0.35  &  150 	&29  & 20.6 \\
		4210  & 25 & 0.061 &  {\color{red} 150}  &27  &49 & &20250 & 30 & 0.35  &  {\color{red} 376}  &29  & 20.6\\
		& & & & & & &26090 & 40 & 0.337 &  125 	&39  & 14.2 \\
		& & & & & & &26090 & 40 & 0.337 &  {\color{red} 418}  &39  & 14.2\\	
	\end{tabular}
	\caption{Details of experimental data; the values highlighted in red indicates $z^+=2.6Re_\tau^{1/2}.$}
	\label{table:exper}
\end{table}

\begin{figure}
	\vspace{-5mm}
	\centering
	\begin{tabular}{cc}
		\subfigure{
			\includegraphics[width=0.71\textwidth] {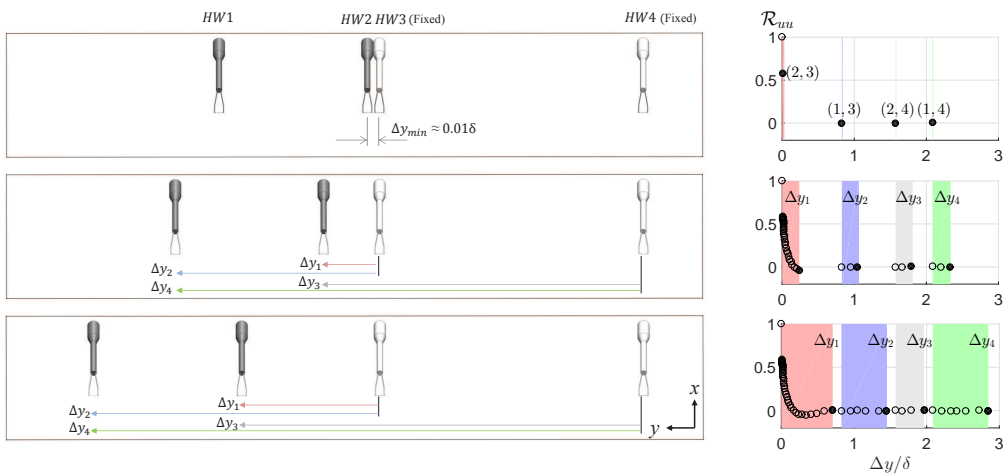}
			\label{expsetupa}
		} & 
		\subfigure{\raisebox{6.5mm}{
				\includegraphics[trim = 0mm 0mm 0mm 0mm, clip, width=3.3cm] {Corr_exp8}
				\label{expsetupb}
		}}
	\end{tabular}
	\vspace{-10mm}
	\caption{Schematic of experimental set-up with four hot-wire probes - HW1, HW2, HW3 and HW4 in the high Reynolds number boundary layer facility and normalized two-point correlation as a function of spanwise separation.}	
	\label{expsetup}
\end{figure}

Measurements of 2-D $u$-spectra were achieved by employing a novel setup that uses two and four single-wire hot-wire probes for the low and high Reynolds number experiments respectively. Hot-wires were operated using an in-house Melbourne University Constant Temperature Anemometer (MUCTA) in all cases. The schematic of the experimental setup for the high Reynolds number measurements is shown in figure \ref{expsetup} (see \citealp{chandran2016} for more details of the low Reynolds number experiment). The setup comprises four hot-wires: HW1, HW2, HW3 and HW4. All hot-wires are calibrated immediately before and after each measurement allowing compensation for drift in the hot-wire voltage during the measurements. The calibration of HW3 is carried out in the freestream with a Pitot-static probe. Since the arrangement did not allow the other three hot-wires to move to the freestream, the calibrated HW3 is used to calibrate HW1, HW2 and HW4, while inside the boundary layer. This is achieved by placing all four wires at the same wall-normal location.  In order to ensure the convergence of mean velocity during the in-boundary layer calibration, the hot-wires are sampled for a longer time, corresponding to $TU_\infty/\delta \approx 10000$. Furthermore, no discernible difference is observed between the 1-D streamwise spectra obtained from the freestream and in-boundary layer calibrated hot-wires. 

Figure \ref{expsetup} schematically shows the arrangement of hot-wires for three sets of spanwise locations and the corresponding calculations of the spanwise correlation. In all cases, HW3 and HW4 are stationed at a fixed spanwise location, whereas HW1 and HW2 are traversed together in the spanwise direction (with logarithmic spacing). Hence the distance between hot-wire pairs, HW1 and HW2 is fixed at all times as is the distance between HW3 and HW4. At the start of the measurement, HW2 and HW3 are as close as practicable to each other with $\Delta y_{min} \approx 0.01\delta$ as the initial spacing between them. The correlation coefficient, $R_{uu}$, as a function of spanwise spacing ($\Delta y$) is obtained by cross-correlating the velocity time series obtained from a pair of hot-wires. Since four hot-wires are employed, each step movement of the traverse gives $R_{uu}$ corresponding to four spanwise spacings: $\Delta y_1$, $\Delta y_2$, $\Delta y_3$ and $\Delta y_4$, which are shown as filled circles in figure \ref{expsetup}. The figure is colour-coded to show the correlation points obtained with a particular pair of hot-wires. The spanwise traversing is carried out up to a maximum spacing of $\Delta y_{max}\approx 2.7\delta$. Therefore, the complete measurement ($\sim 40$ discrete locations) captures $R_{uu}$ for $\Delta y = 0$ and $0.01 \delta < \Delta y < 2.7\delta$, as shown in figure \ref{expsetup}. The use of Taylor's frozen turbulence hypothesis using the local mean velocity as the convection speed allows the construction of correlation functions at different streamwise spacings ($\Delta x$) as well. 
Therefore, the complete 2-D, two-point correlation,  
\begin{equation}
R_{uu}(\Delta x,\Delta y) = \overline{u(x,y)u(x+\Delta x,y+\Delta y)},
\end{equation}
can be calculated (overbar denotes ensemble time-average). A 2-D Fourier transformation of the computed 2-D correlation yields the 2-D spectrum of streamwise velocity fluctuations as a function of $k_x$ and $k_y$,
\begin{equation}
\phi_{uu}(k_x,k_y) =  \int\int_{-\infty}^{\infty}R_{uu}(\Delta x,\Delta y)e^{-j2\pi(k_x\Delta x+k_y\Delta y)} \mathrm{d}(\Delta x) \mathrm{d}(\Delta y),
\end{equation}
where $j$ is a unit imaginary number. 
Here we note that while Taylor's hypothesis is reported to perform well in the outer region for $\lambda_x < 6\delta$ \citep{dennis2008limitations}, uncertainties in the measured spectra are observed for larger scales due to an assumption of a constant convection velocity \citep{del2009estimation,monty2009turbulent}. Recent very large field of view PIV measurements \citep{de2015LDFM} suggest that in the log region, the good agreement between the pre-multiplied spectra from hot-wire and PIV extend to $\sim 15 \delta$ in the streamwise direction. Hence, here we anticipate the effects of Taylor's frozen turbulence hypothesis to be minimal, and do not affect any of the conclusions in this study.

\section{Results and discussions}\label{sec:types_paper}
\subsection{Validation at low Reynolds number}

\begin{figure}
	\centering
	\includegraphics[trim = 0mm 0mm 0mm 0mm, clip, width=10.5cm]{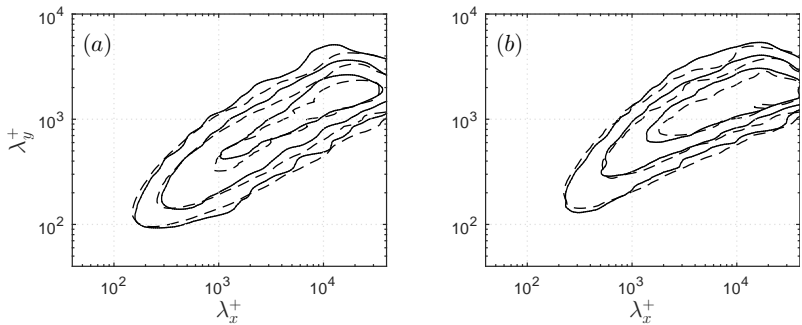}
	\caption{Comparison of corrected experimental 2-D spectra (-----) with DNS (- - -) at $Re_\tau \approx 2000$ for $k_xk_y\phi_{uu}/U_\tau^2 = 0.15,\, 0.25 $ and $ 0.4$ at (a) $z^+\approx100$ and (b) $z^+\approx200$.}
	\label{Valid}
\end{figure}
The measurement technique is validated by comparing the results obtained from the low Reynolds number experiments at $Re_\tau = 2430$ ($\approx 2000$ when computed using $\delta_{99}$) against the boundary layer DNS data of \citet{sillero2014two} at a matched Reynolds number. To this end, a two-dimensional Fourier transformation is carried out on the 2-D correlation obtained from the DNS database. Since the smallest spanwise length scale that can be resolved experimentally with the present arrangement of hot-wires is $\lambda_y^+ = 2\times \Delta y_{min}^+$, a correction scheme based on the available DNS data is employed to account for the unresolved smaller scales. The correction is performed by computing 2-D spectra from DNS data which is interpolated to the experimental spanwise spatial resolution. To this end, the data points in the range $0<\Delta y^+<\Delta y_{min}^+$ are omitted from the original DNS 2-D correlation and later recalculated by linear interpolation in the same way as done for the experiment. The difference between the original and interpolated DNS 2-D spectra corresponds to the amount of energy redistributed due to the insufficient spatial resolution $\Delta y_{min}^+$. The 2-D spectrum calculated from experiment is now corrected by adding this difference to the measured spectrum. The difference is predominant at small scales near $\lambda_y^+ = 2\times \Delta y_{min}^+$ and hence the large scales remain unaffected. Further details of the correction scheme can be found in \cite{chandran2016}.
Figure \ref{Valid} shows the comparison between the constant energy contours of such corrected experimental 2-D spectra and the DNS results, where a good agreement is observed. Further, the correction method is expected to apply at higher Reynolds numbers since inner-scaled energy contributions from small scales are Reynolds number invariant \citep{hutchins2009hot}. Therefore, this correction is applied to all the experimental data presented in this paper for completeness; no conclusions drawn are dependent on the effects of the correction scheme. 

\subsection{Low vs. high Reynolds number spectra}
\begin{figure}
	\centering		
	\includegraphics[trim = 0mm 0mm 0mm 0mm, clip, width=13cm]{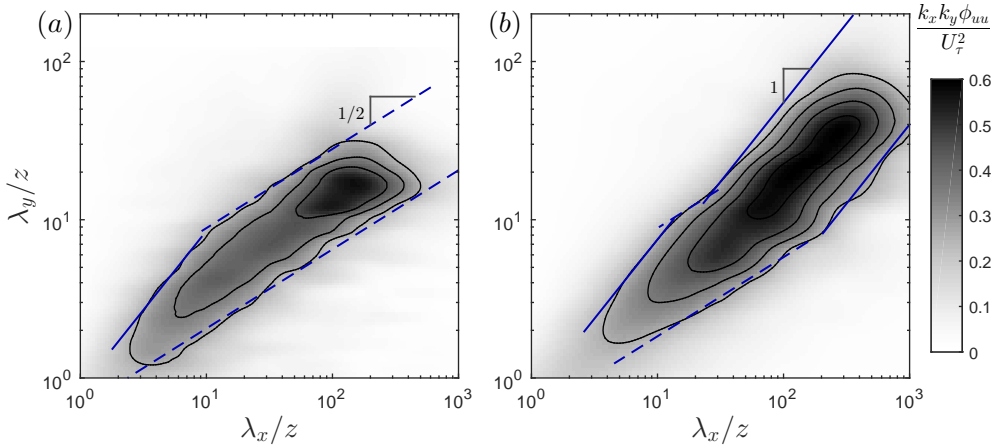}	
	\caption{2-D spectra at  $z^+=2.6 {Re_\tau}^{1/2}$ for (a) $Re_\tau \approx 2400$ and (b) $Re_\tau \approx 26000$; the black line contour represents $k_xk_y\phi_{uu}/U_\tau^2 = 0.25,\, 0.35$ and $0.45$ and the blue solid and dashed lines denote the $\lambda_y/z \sim \lambda_x/z$ and $\lambda_y/z \sim (\lambda_x/z)^{1/2}$ relationships respectively.}
	\label{lowhigh}
\end{figure}

Figure \ref{lowhigh}(a) shows contours of the 2-D spectrum for the $Re_\tau \approx 2400$ case at $z^+ = 116 \, (\approx 2.6Re_\tau^{1/2})$, corresponding to the start of the logarithmic region \citep{klewicki2009logarithmic}. Here, the spectrum is scaled with $U_\tau^2$ and wavelengths are scaled with $z$. The black line contours represent constant energy levels $k_xk_y\phi_{uu}/U_\tau^2 = 0.25, \,0.35$ and $0.45$. A linear relationship of the form $\lambda_y/z \sim \lambda_x/z$ is observed only in the small scale region ($\lambda_x/z,\lambda_y/z < 10$), as marked by the blue solid line. At larger scales, a square-root relationship of the form $\lambda_y/z \sim (\lambda_x/z)^{1/2}$ is evident, as marked by the blue dashed line. These relationships are in agreement with the results of \citet{del2004scaling} from their DNS of turbulent channel flow at a similar Reynolds number. If $\lambda_x$ and $\lambda_y$ characterize the length and width of turbulent structures respectively, such a square-root relationship suggests that the eddies in this scale range are growing longer but not as wide, meaning such eddies are not self-similar.

At high Reynolds number, however, there is a significant difference in the large-scale behaviour. Figure \ref{lowhigh}(b) shows the 2-D spectrum measured at $Re_\tau \approx 26000$ and at $z^+ = 418 \, (= 2.6Re_\tau^{1/2})$. Again, the black line contours represent $k_xk_y\phi_{uu}/U_\tau^2 = 0.25, \,0.35$ and $0.45$ and the blue solid and dashed lines denote the $\lambda_y/z \sim \lambda_x/z$ and $\lambda_y/z \sim (\lambda_x/z)^{1/2}$ relationships respectively. The lower end of the large scale region behaves the same way as in the lower Reynolds number case; i.e, the square-root relationship remains. However, the larger scales deviate away from the square-root relationship and tend towards $\lambda_y/z \sim \lambda_x/z$. This linear relationship, as discussed in \S1, is a necessary condition for self-similarity, which was notably absent for the low Reynolds number case. The length scales at which the peak energy deviates from the low Reynolds number behaviour are approximately $\lambda_x \sim 100z (\sim 1.5\delta)$ and $\lambda_y \sim 15z(\sim 0.2\delta)$. These values denote the minimum dimensions of the scales that are highly energetic and self-similar, and these are certainly large-scales with an average aspect ratio of $\lambda_x/\lambda_y \approx 7$ at the energetic ridge of the spectrum. From an attached eddy perspective, the large aspect ratio observed is consistent with the existence of packets of eddies. Furthermore, the results suggest that the most energetic large-scale structures only become self-similar after maturing into such high aspect ratios.

\begin{figure}
	\centering		
	\includegraphics[trim = 0mm 25mm 0mm 38mm, clip, width=13cm]{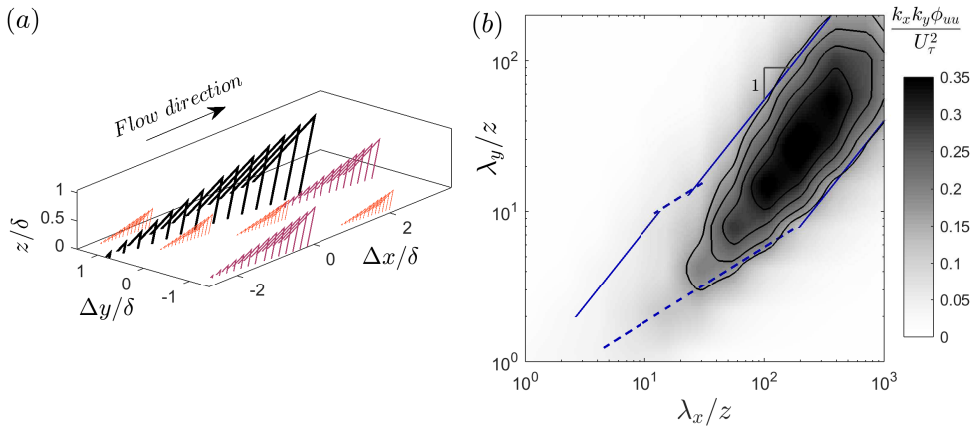}	
	\caption{(a) Illustration of the attached eddy model showing three distinct hierarchies of self-similar packet-eddies with the size of the largest eddy (black) of the order of $\delta$ and (b) 2-D spectrum obtained from attached eddy model at $Re_\tau \approx 26000$ and $z^+=2.6 {Re_\tau}^{1/2}$; blue lines correspond to the constant energy bounds shown in figure \ref{lowhigh}(b).}	
	\label{AttEddy}
\end{figure}

To illustrate the expected qualitative 2-D spectrum from a field of self-similar eddies, here we will utilize the attached eddy model, where hierarchies of representative eddies which are aligned in the streamwise direction, such as those shown in figure \ref{AttEddy}(a), are used to model the logarithmic region \citep{perry1986theoretical}. In figure \ref{AttEddy}(a), for illustrative purposes, a discretized model with distinct hierarchies is shown. However, in the actual simulation, a continuous hierarchy is used with the heights of the smallest and the largest eddies being $100$ viscous units and $\delta$ respectively. Further, the aspect ratio of the representative eddy is roughly the average aspect ratio of the large scales observed in the high Reynolds number results. Figure \ref{AttEddy}(b) shows the 2-D spectrum obtained from this idealized model at similar conditions to the high Reynolds number experiment: $Re_\tau \approx 26000$ and $z^+ = 2.6Re_\tau^{1/2}$. It was shown recently by \citet{baidya2017distance} that asymptotic features such as a $k_x^{-1}$ law are not necessarily observable at finite Reynolds numbers even when self-similarity is strictly enforced. Therefore, it is interesting to note that the contours of the 2-D spectrum at $Re_\tau \sim \mathcal{O}(10^4)$ shown in figure \ref{AttEddy}(b), follow a $\lambda_x \sim \lambda_y$ relation as predicted for the asymptotic state, while the corresponding 1-D spectra still do not exhibit a clear $k^{-1}$ region (see figure \ref{mvsRe}e). Furthermore, the attached eddy results are not expected to follow the square-root relationship observed in experiments (blue dashed line in figure \ref{AttEddy}b) as the model comprises of purely self-similar eddies.

\subsection{A simplified model of 2-D spectra}
\begin{figure}
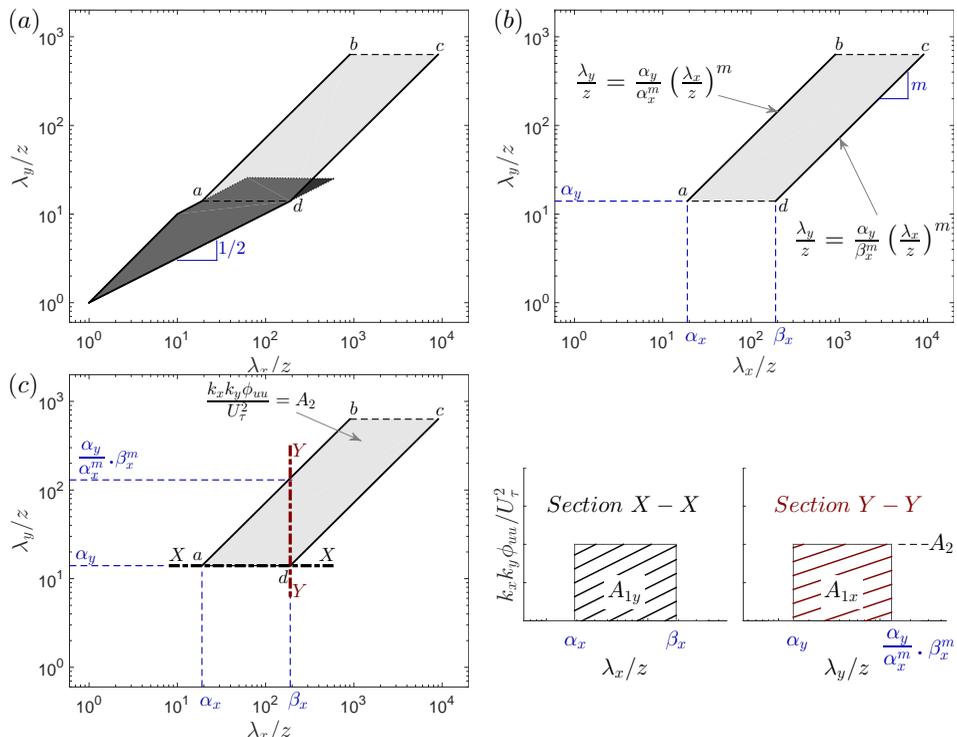

	\centering
	\begin{tabular}{cc}
		\subfigure{
			\label{org1}
			\includegraphics[trim = 0mm 0mm 0mm 2mm, clip, width=6.1cm]{organization_1}} &
		\subfigure{
			\label{org2}
			\includegraphics[trim = 0mm 0mm 0mm 2mm, clip, width=6.1cm]{organization_2}}\\[-3ex]
		\subfigure{
			\label{org3}
			\includegraphics[trim = 0mm 0mm 0mm 2mm, clip, width=6.1cm]{organization_3_2}} &
		\subfigure{\raisebox{8mm}{
				\label{org4}
				\includegraphics[trim = 0mm 0mm 0mm 0mm, clip, width=6.1cm]{organization_4_2}}}
	\end{tabular}
	\vspace{-3mm}
	\caption{(a) Sketch of organization of the 2-D spectra for low (dark shade) and high (light shade) Reynolds numbers, (b) large eddy region (a-b-c-d) with the associated power law and (c) graphical representation of line integrals of the large eddy region with $A_{1x}$ and $A_{1y}$ denoting the respective hatched cross-sectional areas.}
	\label{org}
\end{figure}

Following the sketch of the organization of 2-D spectra at low Reynolds number by \citet{del2004scaling}, let us consider a simplified 2-D spectra model at high Reynolds number for the logarithmic region. The model is idealized by assuming 2-D spectra in the logarithmic region as a constant energy level bounded by the relationships mentioned in figure \ref{lowhigh}. The energy is assumed to be zero outside these bounds. In figure \ref{org1}, the dark shaded patch represents such a region of constant energy for the low Reynolds number case. This is exactly the same as the region sketched in figure 3(c) of \citet{del2004scaling}. The overlaid light shaded patch is the constant energy level for the high Reynolds number case.  Following the experimental results, the large scale region of the high Reynolds number spectra (bounded by a-b-c-d) deviates from the low Reynolds number characteristics and tends towards a higher power law. This region will be referred to as the `Large eddy region'. As described by \citet{chung2015universality}, the upper and lower limits of this region of constant energy are denoted by power laws of the form $\lambda_y/z =  C_1f_1(\lambda_x/z)$ and $\lambda_y/z = C_2f_2(\lambda_x/z)$ respectively. Here ($C_1,C_2$) is the range of the width of eddies for a specified length. As the large eddy region is bounded by parallel lines, this range ($C_1,C_2$) is constant for all eddies of sizes within the large eddy region. Now, if the transition from a square-root relationship occurs at length scales, $\alpha_x$, $\beta_x$ and $\alpha_y$ (as shown in figure 5b), then $(\alpha_x,\beta_x)$ represent the range of {\it lengths} for the transitional eddy {\it width}, $\alpha_y$. In this case, the power laws described by \citet{chung2015universality} could be modified as $\lambda_y/z = \alpha_y/\beta_x^m(\lambda_x/z)^m$ and $\lambda_y/z = \alpha_y/\alpha_x^m(\lambda_x/z)^m$ respectively where the power law coefficient `$m$' is the slope of the large eddy region as shown in the figure. 

In most turbulent flow experiments at very high Reynolds number, only the 1-D spectrum is available. The 1-D spectrum is simply the line integral of the 2-D spectrum at a given wavelength and can be calculated from our model as a function of streamwise and spanwise wavenumbers. If $A_{1x}$ and $A_{1y}$ denote the line integral of the large eddy region (with energy level, say, $A_2$) across the spanwise and streamwise wavenumbers respectively (as shown graphically in figure \ref{org}c), then, 
\begin{subeqnarray}
	A_{1x} =  \frac{k_x\phi_{uu}(k_x)}{U_\tau^2} \bigg | _{\tfrac{\lambda_x}{z} = \beta_x} & = & \int_{\alpha_y}^{\tfrac{\alpha_y}{\alpha_x^m}. \beta_x^m} \frac{k_xk_y\phi_{uu}(k_x,k_y)}{U_\tau^2}\,\mathrm{d}\left(ln \dfrac{\lambda_y}{z}\right) = A_2 ln \frac{\beta_x^m}{\alpha_x^m},\\[5pt] 
	A_{1y} =  \frac{k_y\phi_{uu}(k_y)}{U_\tau^2} \bigg | _{\tfrac{\lambda_y}{z} = \alpha_y} & = & \int_{\alpha_x}^{\beta_x} \frac{k_xk_y\phi_{uu}(k_x,k_y)}{U_\tau^2}\,\mathrm{d}\left(ln \dfrac{\lambda_x}{z}\right) = A_2 ln \frac{\beta_x}{\alpha_x}
	\label{Ax}
\end{subeqnarray} 
and $A_{1x}/A_{1y}=m$. The value of the coefficient, $m$, is therefore the ratio of the constant energy plateaus in the corresponding 1-D streamwise and spanwise pre-multiplied spectra. So when $m=1$, the slope of the bounds of the large eddy region is unity, meaning the eddies within this region have the same energy and they grow proportionally in both the streamwise and spanwise dimensions. Hence, these eddies in the large eddy region satisfy the conditions of self-similarity.
\begin{figure}
	\centering		
	\includegraphics[trim = 0mm 0mm 0mm 0mm, clip, width=11.7cm]{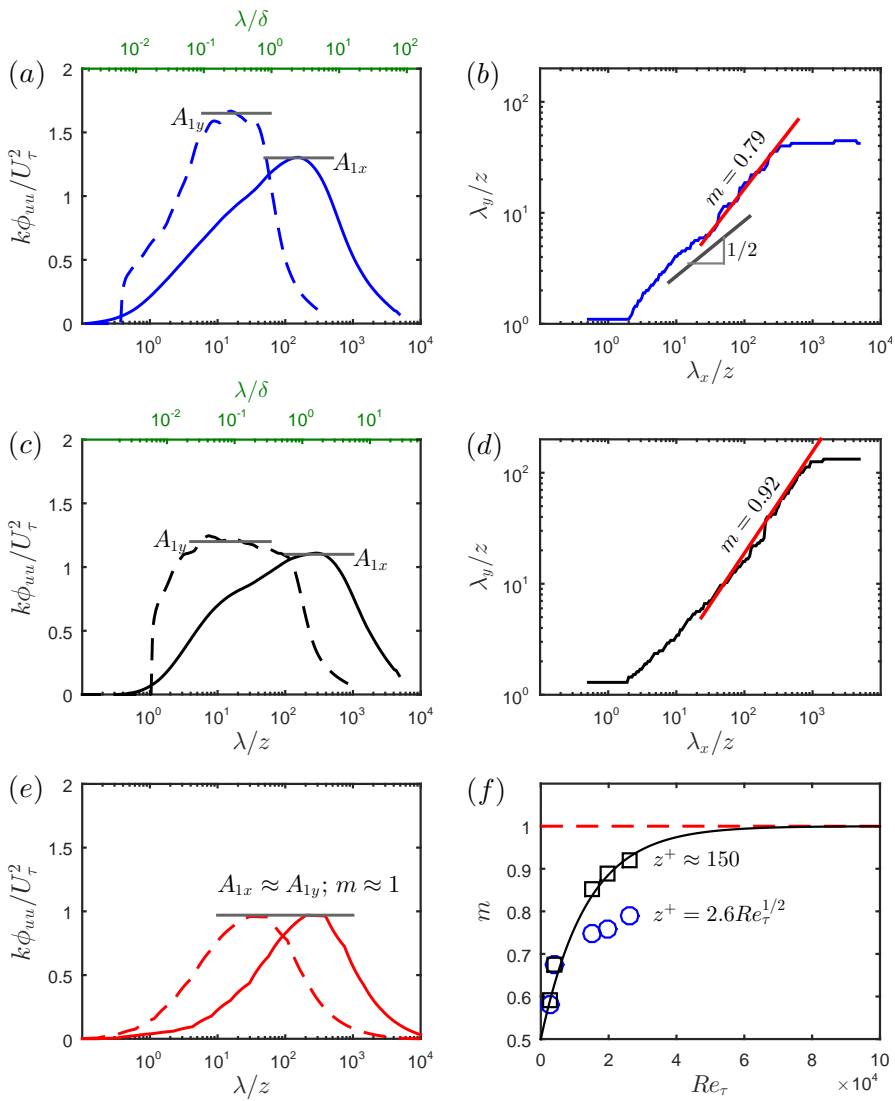}	
	\caption{(a) \& (c) 1-D streamwise (solid) and spanwise (dashed) spectra at $Re_\tau \approx 26000$ for $z^+=418 \, (=2.6 {Re_\tau}^{1/2})$ (blue) and $z^+\approx 150$ (black) respectively, (b) \& (d) ridges of corresponding 2-D spectrum fitted with the calculated slope, (e) 1-D streamwise (solid) and spanwise (dashed) spectra from attached eddy model at matched $Re_\tau \approx 26000$ and $z^+=2.6 {Re_\tau}^{1/2}$ and (f) variation of $m$ versus $Re_\tau$ with the exponential fit of the form $m=1-C_1exp(-Re_\tau/C_2)$.}
	\label{mvsRe}
\end{figure} 

For comparison of the results with the simple spectra model above, figure \ref{mvsRe}(a) shows the 1-D streamwise and spanwise pre-multiplied spectra of $u$ measured at $Re_\tau \approx 26000$ and at $z^+ = 418 \, (=2.6Re_\tau^{1/2})$. Figure \ref{mvsRe}(b) shows the `ridge' of the 2-D spectrum, which is defined as the maximum value of $k_xk_y\phi_{uu}/U_\tau^2$ corresponding to each streamwise wavelength. The deviation of the ridge from $m=0.5$ (square-root relationship) to a higher slope in the large eddy region is evident. The length scales where the 1-D streamwise and spanwise spectra plateau to a constant energy are $\lambda_x \sim 150 z$ ($2.5\delta$) and  $\lambda_y \sim 20 z$ ($0.3\delta$) respectively, and these length scales are within the large eddy region of figure \ref{mvsRe}(b). By considering either the slope of the ridge or the ratio of the peaks of the 1-D energy spectra (figure \ref{mvsRe}a), it is found that $m = A_{1x}/{A_1y} = 0.79$ is a good fit to the data. We now consider the same analysis applied to a lower wall height, $z^+ \approx 150$. Even though this wall location is below the log-region defined by \citet{klewicki2009logarithmic} at this Reynolds number, we might still find some scales of motion exhibiting self-similarity when sufficiently far from the wall. The availability of the 2-D spectra allows us to identify these scales, which would not be possible with only the 1-D spectrum. Figure \ref{mvsRe}(c) shows that the difference between $A_{1x}$ and $A_{1y}$ is less compared to the previous wall-distance. This results in a higher value of $m = 0.92$ (figure \ref{mvsRe}d). The increase in the value of $m$ suggests a greater contribution to the energy at $z^+ \approx 150$ from structures that exhibit a self-similar growth. Provided there are still inertia-dominated motions at this wall-distance, the attached eddy hypothesis predicts a larger population of self-similar eddies to exist as we approach the wall \citep{townsend1980structure} and so the above result is not entirely unexpected. 
Figure \ref{mvsRe}(e) shows the 1-D streamwise and spanwise spectra obtained from the attached eddy model (refer figure \ref{AttEddy}), where only self-similar packets of eddies are present. The results show that $A_{1x} \approx A_{1y}$ such that $m=1$, as expected. However, a clear $k^{-1}$ scaling is still not evident due to an insufficient range of scales within the large eddy region (see figure \ref{AttEddy}b).

The power law coefficient ($m$) is shown above to be an effective indicator of self-similarity and its value can be calculated from both 1-D and 2-D spectra. Arguments for self-similarity require asymptotically large Reynolds numbers, so it would be useful to understand the behaviour of $m$ with increasing Reynolds number. The variation of $m$ for $Re_\tau \approx 2400 - 26000$ is plotted in figure \ref{mvsRe}(f). The value of $m$ increases monotonically with Reynolds number for both the $z^+ = 2.6Re_\tau^{1/2}$ and $z^+ \approx 150$ cases. However, the trend towards self-similarity is more evident at $z^+ \approx 150$. An exponential fit of the form $m=1-C_1exp(-Re_\tau/C_2)$ is reasonable for the $z^+ \approx 150$ case. The value of $C_1$ is taken as 0.5 such that $m \to 0.5$ as $Re_\tau \to 0$, which preserves the square-root relationship of the bounds of high energy in the 2D spectrum at low Reynolds numbers. The value of $C_2$ is simply fitted to the data. Extrapolating the fit to very high Reynolds numbers, it is observed that the value of $m$ approaches unity at $Re_\tau \approx 60000$ for the wall location considered (this is only indicative within the experimental uncertainty of the data). At such a high Reynolds number, a region of constant energy in the 2-D spectra at a wide range of length scales would be found to follow a linear relationship of the form $\lambda_y/z \sim \lambda_x/z$. This means we should not expect to see a clear $k^{-1}$ scaling in both 1-D streamwise and spanwise spectra with $A_{1x} \approx A_{1y}$ for Reynolds numbers below $Re_\tau \approx 60000$. Here we note that the logarithmic law in $\overline{u^2}$ is less sensitive to deviations from self-similarity than the energy spectra \citep{chung2015universality}, and in fact several studies exist where the logarithmic law is observed even when the $k^{-1}$ scaling remained elusive \citep{marusic2013logarithmic}.

\section{Conclusion}
Two-dimensional (2-D) spectra across a decade of Reynolds number ($Re_\tau \approx 2400 - 26000$) for the logarithmic region is presented. While the small-scale contributions are found to be universal when scaled in viscous units, the large-scale contributions show a clear trend for the Reynolds number range examined. Specifically, the contours of 2-D spectra for large streamwise and spanwise wavelengths ($\lambda_x$ and $\lambda_y$) are observed to tend towards a $\lambda_y \sim \lambda_x$ relationship with increasing Reynolds number, starting from a $\lambda_y/z \sim (\lambda_x/z)^{1/2}$ behaviour observed at the lowest Reynolds number measured. It should be noted that the $\lambda_y \sim \lambda_x$ relation indicates self-similarity (i.e. the range of scales with an equal energetic contribution to the streamwise velocity maintain a constant aspect ratio $\lambda_x/\lambda_y$), while the lower Reynolds number $\lambda_y/z \sim (\lambda_x/z)^{1/2}$ behaviour is indicative of structures growing faster in the $x$ direction compared to $y$.

A simple model that describes the 2-D spectral contributions from the large-scales as a region of constant energy bounded by $\lambda_y/z \sim (\lambda_y/z)^m$ is presented. Here the power law coefficient `$m$' corresponds to the slope of the constant energy bounds of 2-D spectra at large scales, or equivalently, the ratio between the constant energy plateaus in 1-D pre-multiplied spectra in the streamwise and spanwise directions. The power law coefficient is proposed to be an effective indicator of self-similarity, and empirical evidence for $m$ monotonically approaching unity with an increase in Reynolds number is presented.

\section*{Acknowledgments}
The authors gratefully acknowledge the financial support of this research from the Australian Research Council.
\bibliographystyle{jfm}
\bibliography{mybibfile}

\end{document}